\documentclass[aps,prl,reprint,showpacs,groupedaddress,superscriptaddress]{revtex4-1} 

\usepackage{multirow} 
\usepackage{url} 
\usepackage{graphicx} 
\usepackage{amsmath,amssymb}

\usepackage{float}
\usepackage[caption=false]{subfig}    % don't load caption or captcont
 
\begin{document} 
 
\title{Some issues concerning the proton charge radius puzzle.} 
  
 \setcounter{figure}{1}
  
\newcommand{\kph}{\affiliation{Institut f\"ur Kernphysik, Johannes  
  Gutenberg-Universit\"at Mainz, 55099 Mainz, Germany.}}

\author{Thomas Walcher}\kph

\date{\today}  
 
\begin{abstract} 
An explanation of the difference of the charge radius of the proton as determined from the Lamb shift in electronic hydrogen and from elastic electron scattering off the proton on the one side and the recent high precision determination with muonic hydrogen on the other side is presented. It is shown that the modification of the $2S_{1/2}$ and $2P_{3/2}$ wave functions by the "Uehling potential"  yields a correction to the theoretical Lamb shift of $\delta (\Delta E_{\textrm{Lamb}} ) = 0.302$\,meV  which has to be compared to $\delta (\Delta E_{\textrm{Lamb}})  = 0.322(46)$\,meV equivalent to the stated radius difference. The explanation is based on the realization that the bound state wave functions modified by the external "Uehling potential" have to be propagated by the vacuum polarization propagator in order to give the corrected leading order Lamb shift. The explanation demonstrates that the Lamb shift is dynamically induced through the QED vacuum polarization and is not only the result of a static external "Uehling potential" probed by a test charge.
\end{abstract}

 
\maketitle  
 
\section{Introduction}
 The seven standard deviations difference of the root-mean-square (rms) charge radius of the proton $r_p = \langle r_p^2 \rangle^{1/2}$ derived from electron scattering and from the Lamb shift of muonic hydrogen has caused a considerable worry in the physics community. Since the Lamb shift is a corner stone of the tests of QED this difference requires indeed a convincing clarification. In this paper a proposal for an explanation in the framework of standard QED is given.
 
 The rms radius derived from the Lamb shift in muonic hydrogen is $r_p  = 0.84184(67)\,$fm \cite{Pohl10} which has to be compared to the CODATA value of 0.8768(69)\,fm \cite{Mohr08} meaning a five standard deviations difference. The CODATA value is derived from a measurement of the Lamb shift in electronic hydrogen \cite{Udem97} and from electron scattering experiments \cite{Sick03}.  Almost at the same time as the muonic experiment a new independent determination with elastic electron scattering has been published and yielded 0.879(8)\,fm \cite{Bernauer:2010wm} resulting in a weighted average of all "electronic experiments" of 0.878(5), i.e. the mentioned seven standard deviations difference. 
 
 In order to transform this radius difference into an energy deviation of the muonic Lamb shift we repeat the key formula for the Lamb shift \cite{Pohl10}:
 \begin{multline}
 \Delta E^{\textrm{theory}}_{\textrm{Lamb}} = (209.9779(49) - 5.2262\,r_p^2 / \textrm{fm}^2 + \\ 
 0.0347\,r_p^3 /\textrm{fm}^3)\,\textrm{meV}
 \label{eq:1}
\end{multline}   
The experiment yielded $\Delta E^{\textrm{exp.}}_{\textrm{Lamb}} = 206.2949(32)\, \textrm{meV}$ from which the radius was determined.
The term with $ r_p^3$ is an approximation to the 3rd Zemach moment $\langle r^3 \rangle_{(2)}$ depending on the charge form factor of the proton. Inserting the differing radii and their error bars given above one gets for the deviation of the Lamb shift:
\begin{equation}
\delta (\Delta E_{\textrm{Lamb}}) = 0.322(46)\,{\textrm{meV}}
\label{eq:2}
\end{equation}
where the error is dominated by the electronic experiments. 

However, a recent non-perturbative relativistic calculation of the theoretical Lamb shift \cite{Carroll:2011rv} yields somewhat different  constants for the formula in eq.\,(\ref{eq:1}).
 From this follows $r_p = 0.83340(67)\,$fm resulting in a deviation $\delta (\Delta E_{\textrm{Lamb}}) = 0.403(51)$\,meV. Considering the perfection of this calculation one is puzzled about possibilities of an explanation. 
 
In view of the excellent accuracy of the muonic experiment and the very good agreement of the about half dozen  electronic experiments, it is highly unlikely that the reason for the deviation is due to a problem on the experimental side. Therefore, one has to see what possibilities are left in the analysis of the data and in the frame work of the QED description. At first one has to realize that the QED calculations of the muonic Lamb shift have been scrutinized again and again over the years. Of the many publications we mention some pertinent summaries \cite{Borie:1982ax,Eides:2000xc,Borie:2004fv,Jentschura:2010ej,Jentschura:2010ha}. On the other hand, there is also no room for a modification of the radiative corrections to the electron scattering cross sections. 

In this paper it is shown that the wave functions determined by Carroll et al. \cite{Carroll:2011rv} have been used incompletely as input in the QED description of the Lamb shift. If correctly used the radius puzzle disappears.
 
\section{Previous proposals for explaining the difference}
 Of the many proposals for explanations we list some prominent and typical.
\begin{itemize}
 \item 3rd Zemach moment\\
 One can construct  an electric form factor with an evanescent charge cloud extending to very large radii but maintaining the radius of the electric experiments. This increases the 3rd Zemach moment $\langle r_p^3 \rangle_{(2)}$ so much that the muonic radius determined from eq.(\ref{eq:1}) agrees with the electric one \cite{DeRujula:2010dp,DeRujula:2010ub,DeRujula:2010zk}. However, it has been shown that such a conjecture is in disagreement with the measured form factors \cite{Cloet:2010qa,Distler:2010zq,Carroll:2011de,Carroll:2011vc}. 
 \item Off-shell form factors\\
 Miller et al.  \cite{Miller:2011yw} have proposed off-shell influences of the form factors in the elastic box diagram contributing to the muonic hydrogen Lamb shift. This idea was refuted by Carlson and Vanderhaeghen who showed that such contributions are two orders of magnitude smaller than thought \cite{Carlson:2011dz}.
 \item Form factor extrapolation problems \\
 The radius is derived from the electric and magnetic form factors $G_{E,M}$ according to $\langle r^2 \rangle^{1/2}=(6/G_{E,M}(Q^2)) (dG_{E,M}(Q^2)/dQ^2)$ at exactly $Q^2 = 0$. Since in elastic electron scattering the form factor can be measured down to small but finite momentum transfers $Q^2$ only one might think about a change of the form factor at very small $Q^2$. Carlson \cite{Carlson:2011zz} refitted a subset of the data of Bernauer \cite{bernauerphd} at very low $Q^2$ but finally got results in agreement with the radius from the complete fits if higher order terms in the form factor $Q^2$ expansion were included. Wu and Kao \cite{Wu:2011jm} tried a "thorn" at very small $Q^2$ equivalent to increasing again the 3rd Zemach moment $\langle r_p^3 \rangle_{(2)}$ so much that agreement from eq.(\ref{eq:1}) results. However, here a dangerous oversight occurs since eq.(\ref{eq:1}) contains the first terms of the perturbative expansion series \cite{Friar:1978wv} only. Since the higher order moments diverge for such extreme modifications of the form factor at very small $Q^2$ eq.(\ref{eq:1}) this expansion breaks down and is not applicable anymore.
    
 Recently the low $Q^2$ extrapolation has been formulated in the realm of chiral dynamics \cite{Ledwig:2011cx} putting it on firm theoretical  grounds  beyond the empirical $Q^2$ expansion. However, no change of the radius is indicated \cite{Vanderhaeghen:2012ab}. Similar ideas have been pursued also in ref.\cite{Pineda:2011xp}.
 
In a very recent paper Lorenz, Hammer, and Mei{\ss}ner  \cite{Lorenz:2012tm} try their own fit to the data of Bernauer et al. \cite{Bernauer:2010wm,bernauerphd} using the "continued fraction" model for the electric and magnetic form factors $G_{E,M}$ \cite{Sick03}. However, of the many models tried by Bernauer et al. this model was one of the worst delivering unstable fits due to poles outside the fitted $Q^2$ range. Only for $Q^2 > 0.1$\,GeV$^2$ the fits of Lorenz, Hammer, and Mei{\ss}ner  are sufficiently stable to allow an extrapolation to $Q^2=0$. However, the poles make such an extrapolation very questionable and consequently also the radius derived \cite{Sick12}. 
    
Also it has to be realized that the good agreement of the electronic Lamb shift and the electron scattering results make explanations with extrapolation modifications unlikely.
\item Dispersion relations\\
In the already cited paper  \cite{Lorenz:2012tm} Lorenz, Hammer, and Mei{\ss}ner  present also a refit of their older dispersion relation fits complementing their old data base with the new data of Bernauer et al. \cite{Bernauer:2010wm}. As the old fits these new fits have a large $\chi^2_{red} =2.2$ for 100 degrees of freedom (dof) estimated from the figure; the dof are not given. Although the use of the theoretical $\chi^2$ distribution is not really justified for these fits since the theory function is unknown (Baysian situation) and the errors are not Gaussian distributed such a $\chi^2_{red}$ is equivalent to a probability $P(\chi^2_{red}  > 2.2 ; dof \approx 100) \approx 10^{-10}$. It also appears that the extraction of the radius from these fits is  "biased", not "consistent" and not "efficient" \cite{Distler:2012mi}. The sensitivity of such fits to the radii is not discussed.
\item Fancy particles beyond the standard model\\
In view of the convincing experimental evidence and the unsuccessful attempts with the mentioned conventional explanations some authors have speculated about new physics. 

Batell, McKeen and Pospelov \cite{Batell:2011qq} consider new vector and scalar forces with carriers of less than 100 MeV mass. Tucker-Smith and Yavin \cite{TuckerSmith:2010ra} propose a new scalar or vector boson of about 1 MeV mass coupling more strongly to the muon than to the electron. Barger et al.  \cite{Barger:2010aj} investigate the possibility of new particles with a special coupling to the muon more generally, but conclude that new spin-0, spin-1 and spin-2 particles are disfavored by other experimental constraints.

In a very recent paper Carlson and Rislow \cite{Carlson:2012pc} discuss two models, one involving new particles with scalar and pseudoscalar couplings, and a second involving new particles with vector and axial couplings. Though it appears not impossible to accommodate the new particles, some fine tuning of masses and couplings is needed to adjust them to the used constraints of the Lamb shift, muon magnetic moments and kaon decay rate data.
\end{itemize}

\section{An overlooked issue}
The muonic Lamb shift is in leading order caused by the vacuum polarization in QED described by the exchange of an electron-positron loop. The loop correction can be calculated perturbatively with Feynman diagrams, see e.g. ref.\,\cite{Weinberg:1995mt,Pachucki:1996zza}, or non perturbatively from the exact solution of the Dirac equation using Green's function \cite{Wichmann:1956zz,Gyulassy:1974ba}. Not surprisingly both approximations give the same result in leading order of the expansion in $Z  \alpha^2$. We use the first approach since it is the right framework for radiative corrections \cite{Weinberg:1995mt}. 

Using the notation of Weinberg \cite{Weinberg:1995mt} $u_B$ is the unperturbed solution of the wave equation of the muon in the external Coulomb field:
\begin{equation}
 H_0 u_B  = E^{(0)}_B  u_B  \;  ;  \; H_0 = T+ V_{\textrm{Coulomb}}
\label{eq:3}
\end{equation}
The energy shift is then given in leading order by the following expression:
\begin{equation}
E^{(0)}_{\textrm{Lamb},B} = -\langle u_B |\Sigma^*_{\cal{A}} | u_B \rangle.
\label{eq:4}
\end{equation}
where  $\Sigma^*_{\cal{A}} $ the "self energy function". It comprises in principle the self energy, the vertex correction and the vacuum polarization. The formula in eq.(\ref{eq:4}) is derived from the propagator of the bound muon and a first order expansion in energy with respect to the unperturbed energy eigenvalue $E^{(0)}_B$ (\cite{Weinberg:1995mt}, chapter 14.2).  Since the matrix element is of the form $\int \Sigma^*_{\cal{A}}  \rho(r) r^2 dr$ with $\rho(r) = |u_B|^2$ the charge distribution of the muon orbit, it is suggestive of an external  potential $V_{VP} =-\Sigma^*_{\cal{A}}$, the "Uehling potential" $V_{\textrm{Uehling}}  = V_{VP}$. However, this identification is not a trivial step, but rather an approximation which has to be justified in the realm of QED. 

In this paper we want to investigate this approximation and examine higher order effects. The first part of this examination is based on a note of Carl Carlson \cite{Carlson:2012ab} extending the leading order derivation of Weinberg to all orders. However, we shall see that even this is not enough. In diagrammatic form we have a generalized Dyson series as depicted in Fig.\,\ref{fig:VP_4}(a):
 
\begin{figure}[H]
\centering
\subfloat[expanded form]{\includegraphics[width=0.7\columnwidth]{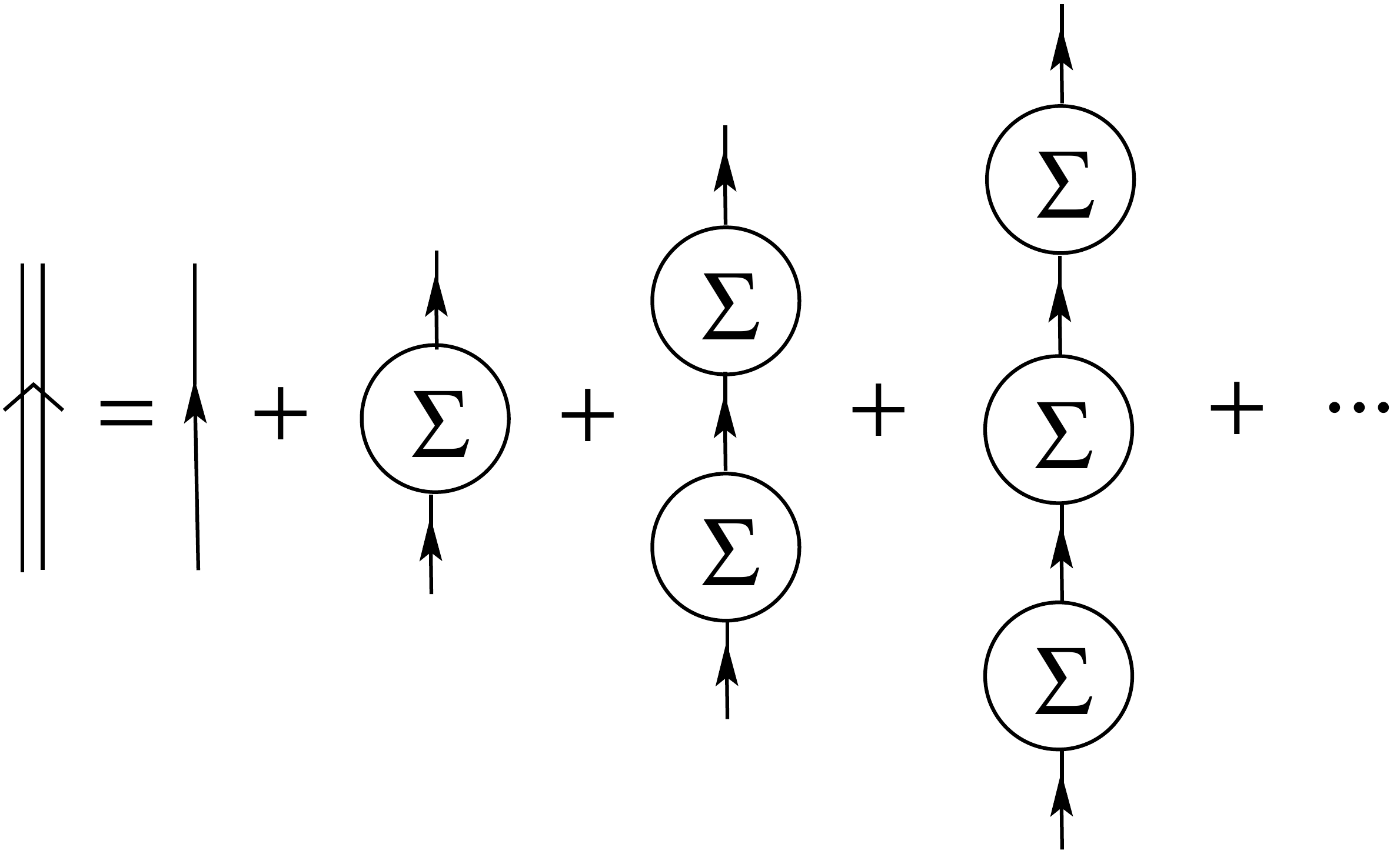}} 
\end{figure}
\begin{figure}[H]
\ContinuedFloat
\centering
\subfloat[reiterated form]{\includegraphics[width=0.5\columnwidth]{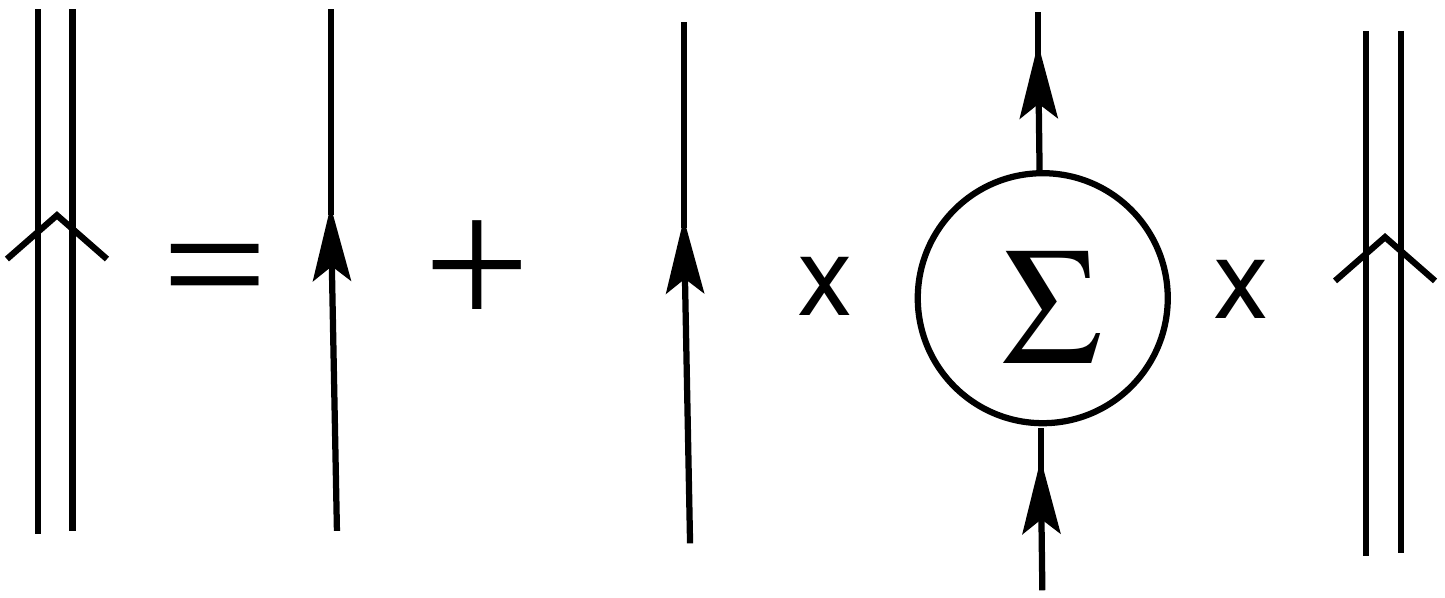}}
\caption{The generalized Dyson series in the presence of the external Coulomb potential. The full lines represent the muon propagator  $S_{\cal{A}}$ with the Coulomb wave functions $u_B$, the double full line the propagator $S'_{\cal{A}}$ with the wave function $U'_B$ modified by the higher order terms.}
\label{fig:VP_4}
\end{figure}

Let $S_{\cal{A}}$ be the propagator of the muon which contain the binding by the Coulomb potential and $S'_{\cal{A}}$ the propagator including the additional effect of the electron-positron loops, as shown in Fig.\,\ref{fig:VP_4}. We get the Dyson equation generalized for bound states:
\begin{equation}
S'_{\cal{A}} = S_{\cal{A}} + S_{\cal{A}} \Sigma^*_{\cal{A}}  S'_{\cal{A}}
\label{eq:5}
\end{equation}
or, after Fourier transformation from time to energy and making the integrals explicit we get in Weinberg's notation:
\begin{multline}
S'_{\cal{A}}(x,y,E) = S_{\cal{A}}(x,y;E) + \\  \int d^3z \int d^3 w \, S_{\cal{A}}(x,z;E) \Sigma^*_{\cal{A}}(z,w;E)  S'_{\cal{A}}(w,y;E)
\label{eq:6}
\end{multline}
where  $u_B$ is the unperturbed wave functions and $U'_B$ the wave function perturbed by the vacuum polarization potential $V_{VP}$. In the usual approximation this is the external "Uehling potential", but we shall have to come back to this point. 
\begin{equation}
H U'_B  = E'_B  U'_B \;  ;   \; H = T+ V_{\textrm{Coulomb}} + V_{VP}
\label{eq:7}
\end{equation}
(For reasons which will become clear later we have deviated from Weinberg's notation and added a $'$ to the $U_B$.)
With this one has:
\begin{eqnarray}
S_{\cal{A}}(x,y,E) = \sum_N \frac{u_N(x)\bar{u}_n(y)}{E_N-E-i \epsilon} \label{eq:8} \\
S'_{\cal{A}}(x,y,E) = \sum_N \frac{U'_N(x)\bar{U}'_N(y)}{E'_N-E-i \epsilon} \label{eq:9}
\end{eqnarray}
where we have omitted the terms with negative energy states. $u_N$ and $U'_N$ are normalized to one. One can now evaluate eq.(\ref{eq:5}) by multiplying from the right with $\gamma^0 U'_B(y)$ and integrate over $y$, then multiply from the left with $u_B^{\dagger}(x)$ and integrate over $x$, and then solve for $\delta E'_B = E'_B - E^{(0)}_B$ yielding \cite{Carlson:2012ab} :
\begin{equation}
E^{(1)}_{\textrm{Lamb},B} = \delta E'_B = -\frac{\int d^3x\, \int d^3y \, \bar{u}_B(x) \Sigma^*_{\cal{A}}(x,y;E) U'_B(y)}{\int d^3x\, u_B^{\dagger}(x) U'_B(x)}
\label{eq:10} 
\end{equation}
where the (1) indicates that we have now included all next to leading order diagrams. If we replace the perturbed $U'_B(y)$ by the unperturbed $u_B(y)$ we regain eq.(\ref{eq:4}). The normalization factor $\int d^3x\, u_B^{\dagger}(x) U'_B(x)$ contributes to third order in $Z \alpha^2$  and will not be considered further, however, it is included in the calculation presented beneath.

\begin{figure}[H]
\centering
\includegraphics[width=0.4\columnwidth]{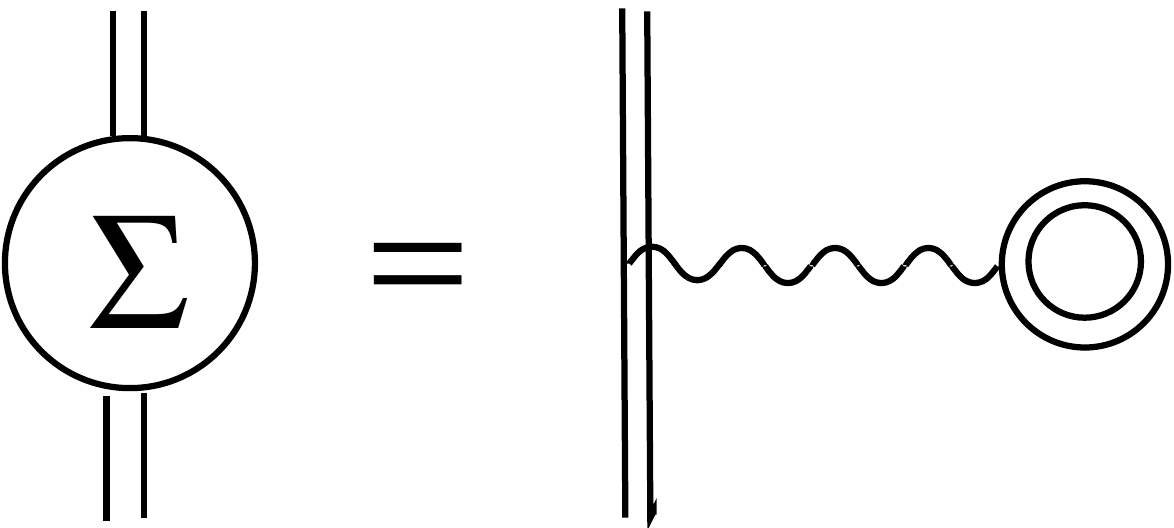}
\caption{The correction term $\delta S_{\cal{A}}(x,y;E)$ to the unperturbed propagator  $S_{\cal{A}}(x,y;E)$ due to the muon self energy function $\Sigma^*_{\cal{A}}(x,y)$ in the presence of an external potential. The double straight lines indicate the muon propagator with the stationary wave function $U_B$ which comprises the influence of the external Coulomb field plus the vacuum polarization field. The double loop line indicates the electron-positron loop in the external Coulomb field.}
\label{fig:VP_1}
\end{figure}

The previous considerations neglect, however, two facts. Firstly, the initial and final propagator, i.e. wave functions, should be the same for a stationary bound state. Otherwise only a transient state with a lifetime of $\delta t = \hbar / \delta E'_B$ is described.  Therefore, the propagator in Fig.\,\ref{fig:VP_1}  for the vacuum polarization effect should be "dressed" by the unknown perturbed stationary wave function $U_B$. Since the vacuum polarization contribution depends on the wave functions, the $U_B$  has to be calculated self-consistently from eq.(\ref{eq:7}). Secondly, the self energy function the electron-positron loop is not the bare one, since it it has to include the exchange of photons and integration over the associated internal momenta. Therefore, it is not possible to show the equivalence of the one-loop electron-positron exchange to the sum of the ladder of these exchanges, as Weinberg shows for the one photon exchange and the Coulomb potential (\cite{Weinberg:1995mt}, chapter 13.6). As Weinberg states all corrections of higher order may be included as radiative corrections to the leading external Coulomb potential.

The correction term in Fig.\,\ref{fig:VP_1}  is unknown and one has to find a reasonable approximation to calculate it. The external "Uehling potential" derived from for the matrix element of the undressed self-energy function with the unperturbed wave function eq.(\ref{eq:4}) is a leading order approximation only. If we solve the wave equation eq.(\ref{eq:7}) with the external "Uehling potential" all contributions due to higher order loop exchanges are missing in the wave function.

Actually we have to solve the two equations eq.(\ref{eq:5}) and eq.(\ref{eq:7}) simultaneously. The usual method to do this is by iteration. We first determine $u_B$ from eq.(\ref{eq:3}), use the bare propagator in eq.(\ref{eq:8}) and then arrive at eq.(\ref{eq:4}) as Weinberg. From this we get the leading order approximation $V_{VP}$, i.e. the "Uehling potential". With this we solve eq.(\ref{eq:7}), get the propagator in eq.(\ref{eq:9}) insert it in eq.(\ref{eq:5}) and get a new  $V'_{VP}$. In diagrammatic form this looks like:

\begin{figure}[H]
\centering
\includegraphics[width=0.4\columnwidth]{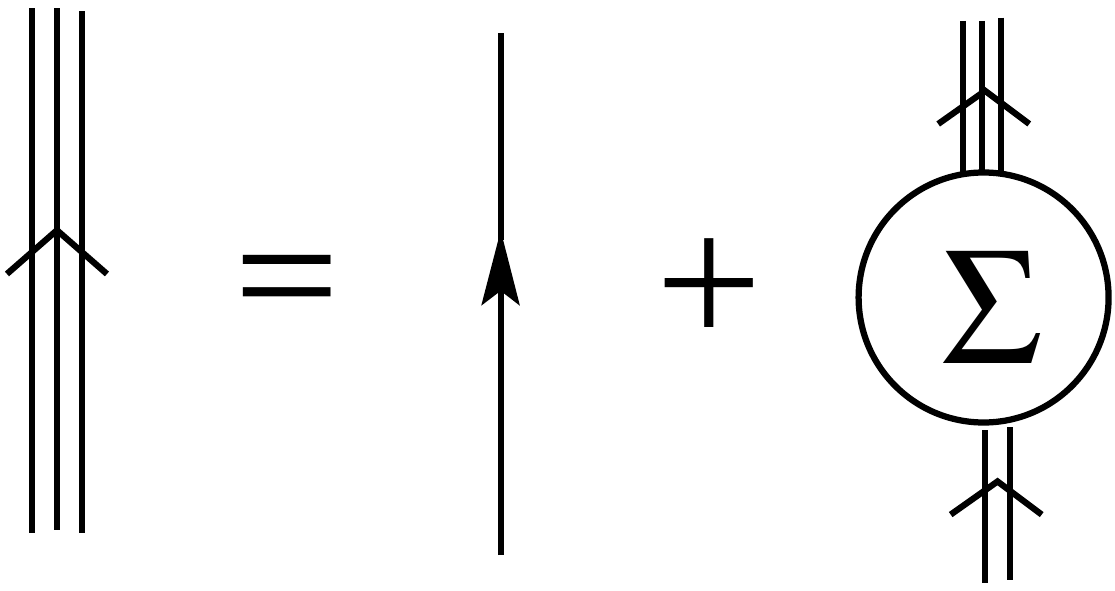}
\caption{One iteration of the Dyson equation generalized to the presence of the external Coulomb field plus the field due to the  vacuum polarization in diagrammatic form.}.
\label{fig:VP_5}
\end{figure}

The continued procedure can be written as:

\begin{alignat}{2}
&H_0 u_B  =  E_B^{(0)} u_B ;  & H_0 & = T + V_{\textrm{Coulomb}} \\
&S'_{\cal{A}}   =  S_{\cal{A}} + S_{\cal{A}} \Sigma^*_{\cal{A}}  S_{\cal{A}} & & \longrightarrow   V_{VP}   \\
&H' U'_B  = E'_B U'_B ; & H' & = H_0 + V_{VP} \\
&S'_{\cal{A}}   =  S_{\cal{A}} + S_{\cal{A}} \Sigma^*_{\cal{A}}  S'_{\cal{A}}  & & \longrightarrow   V'_{VP} \\
&H'' U''_B  = E''_B U''_B ; & H'' &= H_0 + V'_{VP} \\
&S''_{\cal{A}}  = S_{\cal{A}} + S'_{\cal{A}} \Sigma^*_{\cal{A}}  S''_{\cal{A}}  & & \longrightarrow   V''_{VP}  \label{eq:16} \\
& \ldots \nonumber
\end{alignat}
This procedure is known from many body physics where the self energy function is modified by many-body feed back effects or an external potential \cite{Mattuck:1976xt}. Here the muon propagator is modified by the vacuum polarization effect. The iteration can be viewed as a "self consistent time-dependent approximation".  A physical picture of this procedure is that the wave functions are reiterated until the propagators stop changing and the calculation becomes "self-consistent". This means that the state $U^{(n \prime)}_B $ into which the state $U^{((n-1) \prime)}_B $ is "scattered" lives longer and longer. The stationary state is of course only approached, reflecting the fact that a propagator cannot describe a stationary bound state \cite{Newton:1982qc}. In the limit the self-energy function transfers only momentum and no energy, in accord with the idea of a particle in a stationary state circling in a central potential. 

We can follow a similar derivation which yielded eq.(\ref{eq:10}) now for eq.(\ref{eq:16}) and get:
\begin{multline}
 E''_B - E^{(0)}_B = \\
 -\frac{\int d^3x\, \int d^3y \,\bar{U'_B(x)} \Sigma^*_{\cal{A}}(x,y,E) U''_B(y)}{\int d^3x\, U_B^{\prime \dagger} U''_B(x)} \\
 + {\cal{O}} \left(\frac{\delta E^{\prime 2}_B}{(E_B-E)} \right)
\label{eq:17}
\end{multline}

If we approximate $U''_N \approx U'_N$, i.e. the wave function calculated with the Uehling approximation and neglect the higher order term as Weinberg and the normalization denominator $\int d^3x\, U_B^{\prime \dagger} U''_B(x)$ , we get the result for the approximately stationary vacuum polarization propagator:
\begin{multline}
E^{(2)}_{\textrm{Lamb},n\,l} = E''_B - E^{(0)}_B = \\
 -\int d^3x\, \int d^3y \,\bar{U}'_B(x) \Sigma^*_{\cal{A}}(x,y,E) U'_B(y)
\label{eq:18}
\end{multline}
The result is also intuitively satisfying since one expects that the wave functions are not only determined by the external Coulomb potential but also by the external "Uehling potential". There is a feedback effect of the vacuum polarization which can be made transparent iteratively. 

It is not evident from eq.(\ref{eq:18}) that the "Uehling potential" mixes the intermediate excited states. This can be seen from the solution to second order in time independent perturbation theory where we use the usual bra-ket notation \cite{Sakurai:2011zz}:
\begin{equation}
E_B-E_B^{(0)} = \langle u_B |V_{VP} | u_B\rangle + \sum_{B \neq M} \frac{| \langle u_B |V_{VP} | u_M  \rangle |^2 }{E_B^{(0)} -E_M^{(0)} }
\label{eq:19}
\end{equation}
The second term represents the mixing of, or "scattering into", the intermediate states $M$ and is of order $(Z \alpha^2)^2$. There is no contribution diagonal in the unperturbed wave function $u_B$ in this order which would belong to the "double vacuum polarization" \cite{Pachucki:1996zza} or "vacuum polarization iteration" \cite{Borie:2004fv} or "polarization insertion in two Coulomb lines" of Carroll et al. \cite{Carroll:2011rv}. The reason is simply that there is no part due to this diagram in the vacuum polarization potential, only the "Uehling potential" is present in the Hamiltonian. This means that the contribution of this diagram is missing in the wave functions and if we use them to calculate the energy shift with eq.(\ref{eq:18}) we miss it. On the other hand this contribution is present if we would calculate the series in Fig.\,\ref{fig:VP_4}(a) with the unperturbed wave functions $u_B$. Therefore, the "double vacuum polarization" contribution has to be calculated separately and we have taken the value of Pachucki \cite{Pachucki:1996zza}.

At this point it has become clear that the usual use of the external "Uehling potential" \cite{Uehling:1935uj} for the vacuum polarization effect is an approximation only. This point is essential because it means that one assumes that the muon can be regarded as a test charge in the external "Uehling potential"  \footnote{The problem of the test charge in QED played a key role in the development of QED \cite{Feynman:1965er}.}  and neglects the QED feedback effects. 

If one solves eq.(\ref{eq:7}) numerically the eigenvalue is exact, or "non-perturbative" as it is called by Carroll et al. \cite{Carroll:2011rv}, in the realm of time independent perturbation theory, but it is the "perturbative" approximation discussed above in time dependent perturbation theory. Consequently, we call the energies and wave functions calculated with $V_{VP}$ approximated by the "Uehling potential" $V_{\textrm{Uehling}}$ "perturbed" and without "unperturbed". 

We can derive from eq.(\ref{eq:7}) a form corresponding to eq.(\ref{eq:10}).
\begin{equation}
E^{(1)}_{\textrm{Lamb},B} = E_B - E^{(0)}_B  = \frac{\langle u_B |V_{\textrm{Uehling}} | U'_B \rangle}{\langle u_b| U'_B \rangle}
\label{eq:20}
\end{equation}
This equation is  identical to eq.(\ref{eq:10}) but differs from eq.(\ref{eq:18}). This means that the time independent eigenvalue equation cannot describe the dynamical feedback effects of QED which become visible through the iteration presented above. This is conceptional not surprising since the lowest order Lamb shift is a pure QED effect not present in the eigenvalue of the time independent wave equation. The next to leading order approximation with the external "Uehling potential" is not enough after the precision of the experiments has been improved to the level of Pohl et al. \cite{Pohl10}.

It is not correct to assume that the "polarization insertion in two Coulomb lines", Carroll et al. \cite{Carroll:2011rv}, is already present in the solution of the time independent wave equation eq.\,(\ref{eq:7}). Rather this contribution has to be calculated separately applying Feynman rules and added to the propagator for $\Sigma^*_{\cal{A}}$ as indicated in Fig.\,\ref{fig:VP_4}(a). This is the "double vacuum polarization" calculated by Pachucki \cite{Pachucki:1996zza}.
What is present in the "non-perturbative" solution of Carroll et al. is the scattering into the intermediate excited states. The expression eq.(\ref{eq:18}) is intuitively plausible and has been already mentioned by Carroll et al. \cite{Carroll:2011rv}, but, was not used and discussed by them.

\section{Numerical realization \label{sec:nr}}
In distinction to Carroll et al. in ref.\,\cite{Carroll:2011rv} we restrict ourselves to the non relativistic Schr\"odinger equation with the Coulomb potential of a point charge. This suffices for demonstrating the correction of the Lamb shift without considering fine structure splitting, finite size effect, etc..  Relativistic effects on the Lamb shift are small \cite{Borie:2004fv} and do not change anything essential for this discussion. This means we calculate the exact solutions of the Schr\"odinger equation with the point Coulomb potential and the external "Uehling potential" added according to  eq.(\ref{eq:5}).

A compact representation of the "Uehling potential" $V_{\textrm{Uehling}}$, well suited for our calculation, is the representation provided by Pachucki \cite{Pachucki:1996zza}: 
\begin{equation}
V_{\textrm{Uehling}} (r) = -\frac{Z \alpha}{r} \frac{\alpha}{\pi} \int_4^{\infty} \frac{d(q^2)}{q^2} \exp(- m_e q r) \,u(q^2)
\label{eq:21}
\end{equation}
where
\begin{equation}
u(q^2) = \frac{1}{3} \sqrt{\left (1-\frac{4}{q^2} \right )} \left (1+ \frac{2}{q^2} \right )
\label{eq:22}
\end{equation}
and $q^2$ is the internal momentum squared normalized to $m_e^2$. For the radial unperturbed wave functions we use: 
\begin{eqnarray}
R_{2S} &=& \frac{1}{\sqrt{2}} \frac{1}{\sqrt{a_0^3}} \exp(-\frac{r}{2 a_0}) (1- \frac{r}{2 a_0})  \label{eq:23} \\
R_{2P} &=& \frac{1}{2 \sqrt{6}} \frac{1}{\sqrt{a_0^3}} \exp(-\frac{r}{2 a_0} ) \frac{r}{a_0} \label{eq:24}
\end{eqnarray}
with $a_0 = \hbar c/(\alpha \mu)$ the Bohr radius and $\mu$ the reduced mass. Using these wave functions in eq.(\ref{eq:4}) we get for the leading order Lamb shift $\Delta E_{Lamb} = E^{(0)}_{\textrm{Lamb},2\,P} - E^{(0)}_{\textrm{Lamb},2\,S}  = 205.005$\,meV in agreement with \cite{Pachucki:1996zza}.

For the numerical integration of the Schr\"odinger equation we have used Mathematica. As Carroll et al. \cite{Carroll:2011rv} we have made extensive tests to guaranty the quality of the solutions. All calculations have been made with an internal precision of 64 digits and an accuracy goal of 20 digits. The optimal method is the "Explicit Runge-Kutta" integration for the S-State and the change between various methods provided "automatically" for the P-state. The numerical eigenvalues of the unperturbed $2S$ and $2P$ states, i.e. without the "Uehling potential", are compared to the non relativistic exact solution (Bohr energies) and found to be good to a few neVs for different boundary conditions at small ($\approx 0.1$\,fm) and large ($\approx 10000$\,fm) radii to which the eigenvalues are sensitive. Of course, since we do not take the difference of the large energy eigenvalues as Carroll et al. \cite{Carroll:2011rv} but calculate the small Lamb shift only, we do not really need this extreme accuracy. However, when using our unperturbed numerical eigenvalues we get $\Delta E_{Lamb}  = 205.005$\,meV in complete agreement with the calculation using exact wave functions of eq.(\ref{eq:23}) and eq.(\ref{eq:24}). The eigenvalues with the "Uehling potential" have been determined using the virial theorem.

\begin{figure}[h]
\centering
\includegraphics[width=0.95\columnwidth]{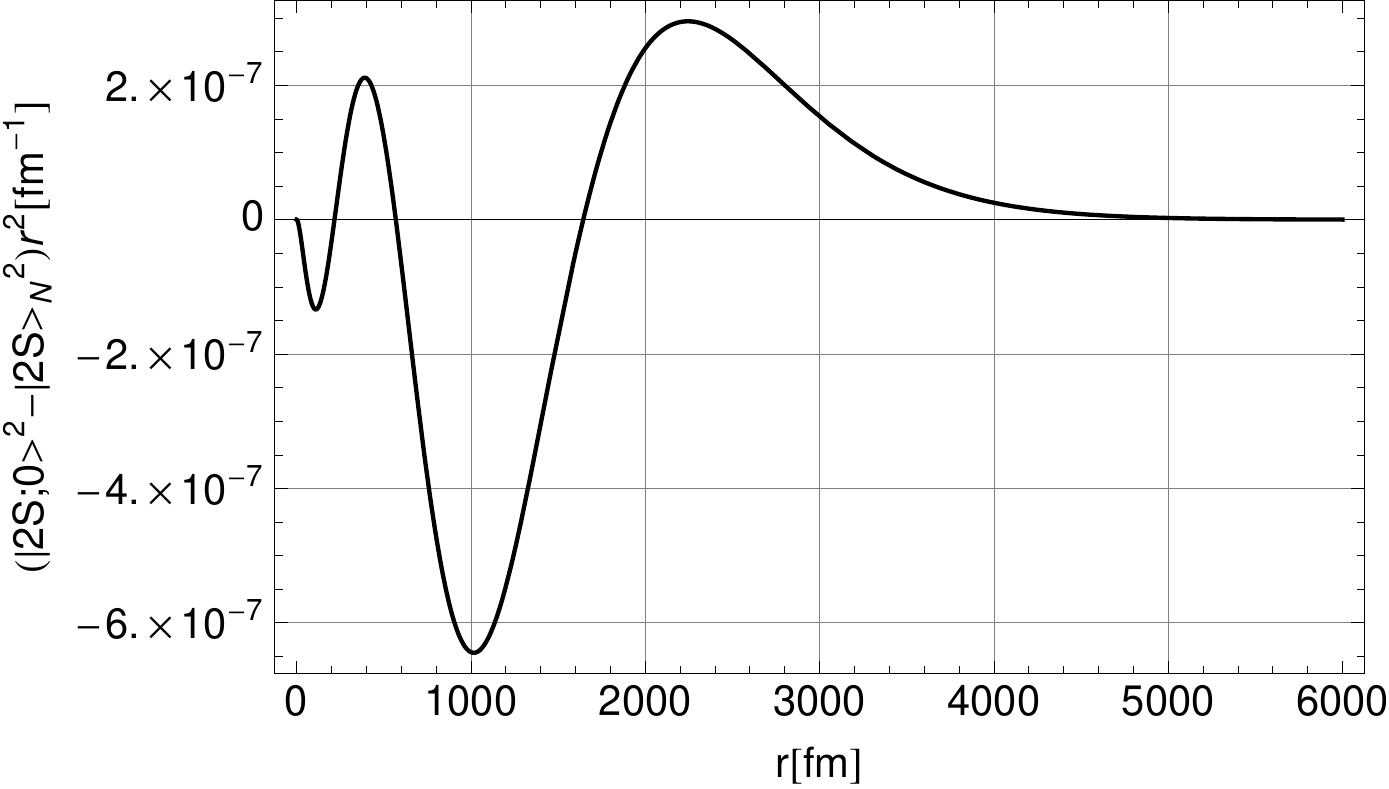}
\caption{The difference $(| 2\,S; 0 \rangle^2 -  | 2\,S \rangle_N^2) r^2$ showing the polarization charge density divided by the negative elementary charge of the muon due to the "Uehling potential" for the $2\,S$ state}.
\label{fig:1}
\end{figure}

\begin{figure}[h]
\centering
\includegraphics[width=0.95\columnwidth]{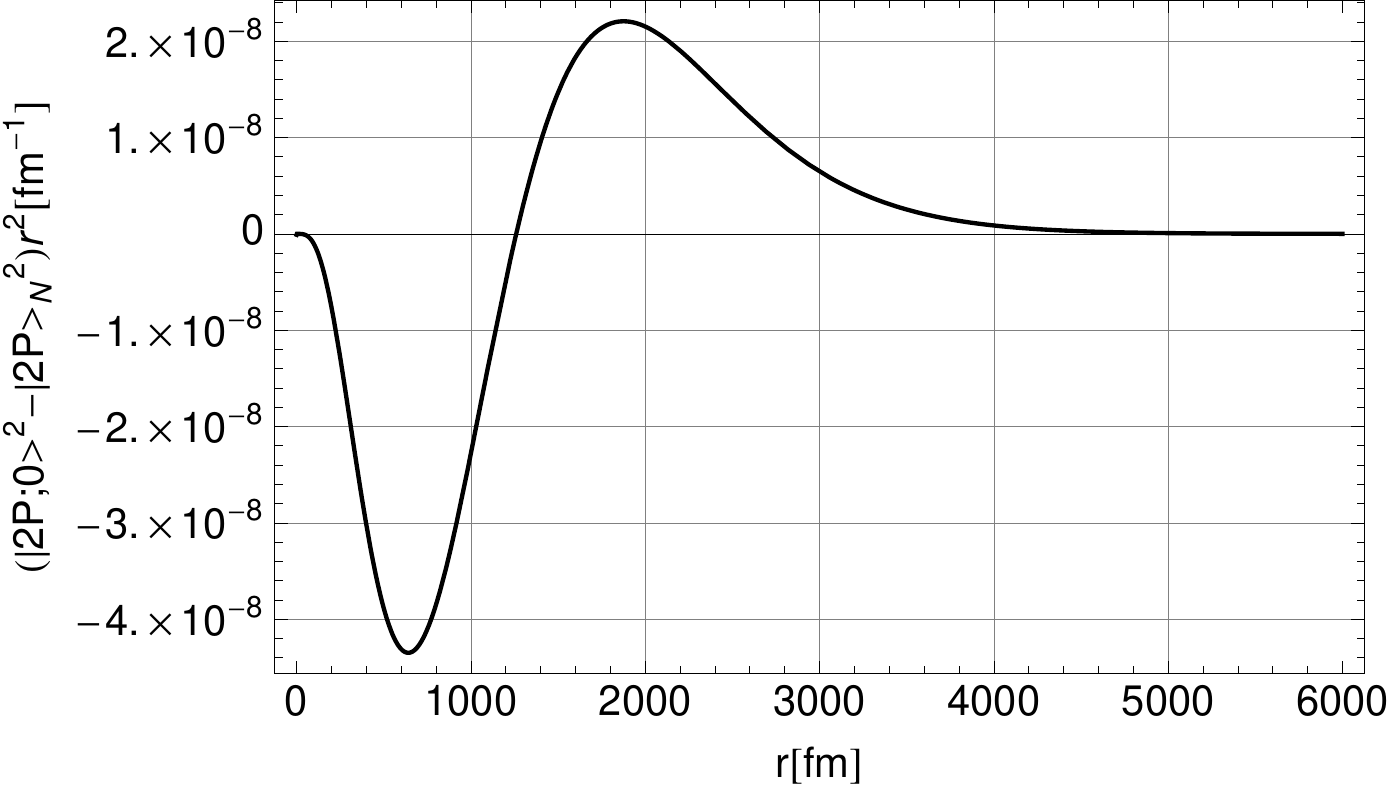}
\caption{The difference $ (| 2\,P; 0 \rangle^2 -  | 2\,P \rangle_N^2) r^2$ showing the polarization charge density divided by the negative elementary charge of the muon due to the "Uehling potential" for the $2\,P$ state}.
\label{fig:2}
\end{figure}

Figure \ref{fig:1} and Fig. \ref{fig:2} show the difference of the density of the unperturbed state $| n\,l ; 0 \rangle^2 r^2 = u_B^2 r^2$ minus the density of the normalized perturbed state $| n\,l  \rangle_N^2 r^2 = U{\prime\, 2}_B r^2$. 

One nicely sees the polarization charge induced in the vacuum by the muon. The wiggles at small radii in Fig. \ref{fig:1} are no numerical artifacts but due to the double bump structure of the 2S state.  In order to get an idea of the scales we note that the Bohr radius for the muon is $a_0=285$\,fm, the scale of the Compton wave length of the electron or positron in the electron-positron pair of the vacuum polarization $\lambda = 2426$\,fm, the rms radius of the 2S state $\langle2\,S | r^2 |2\,S \rangle^{1/2} = 1854$\,fm, and for the 2P state  $\langle2\,P | r^2 | 2\,P\rangle^{1/2} = 1560$\,fm. As expected some positive charge is pushed to larger radii compensated by negative charge at small radii indicating the induction of the polarization cloud in the vacuum. An analytical derivation of this polarization is given in ref.\,\cite{Weinberg:1995mt}, chapter 11.2.

Calculating the difference of the Lamb shifts with the normalized perturbed $2S$ and $2P$ states according to  eq.(\ref{eq:18}) one gets the salient result of this paper:
\begin{equation}
\Delta E^{\textrm{point charge}}_{\text{Lamb}} = 205.307(1) \textrm{meV}
\label{eq:25}
\end{equation}
where the error is a best estimate from the variation of the value with different integration boundaries. Comparing this result to the canonical value for the unperturbed wave functions $\Delta E^{(0),\textrm{point charge}}_{\text{Lamb}} = 205.005(1)$ one arrives at 
\begin{equation}
\delta (\Delta E_{\textrm{Lamb}})  = 0.302(1)\,\textrm{meV}
\label{eq:26}
\end{equation}
in very good agreement with the searched for difference of eq.\,(\ref{eq:2}).  

For the difference of the eigenvalues from the numerical solution of the wave equation according to eq.(\ref{eq:20}) we get $E_{2\,P} -E_{2\,S} = 205.156(1)\,\textrm{meV}$. If we take directly the eigenvalues $E_{2\,P} -E_{2\,S}$ from the numerical solution of the wave equation eq.(\ref{eq:7}) as Carroll et al. \cite{Carroll:2011rv} we get $205.159(3)\,\textrm{meV}$. The deviation from the difference calculated with eq.(\ref{eq:20}) indicates a limit of the numerical accuracy. Since we do not use the difference of the eigenvalues we have not insisted to improve this limit. Considering the relativistic correction due to the Dirac wave functions of $0.021\,\textrm{meV}$ \cite{Pachucki:1996zza}, missing in our non relativistic calculation, this is in good agreement with the relativistic result of Carroll et al. \cite{Carroll:2011rv} of $205.1706(5)\,\textrm{meV}$. 

As already discussed the result in eq.(\ref{eq:25}), does contain the contribution due to the scattering into intermediate excited states. To this the "double vacuum polarization" contribution of 0.151\,meV of Pachucki \cite{Pachucki:1996zza} has to be added. Carroll et al. \cite{Carroll:2011rv} calculated the Lamb shift from the Dirac equation yielding a value equivalent in the non relativistic limit to eq.(\ref{eq:10}) and to eq.(\ref{eq:19}) up to order $Z^2 \alpha^4$. This yields effectively only half of the value given in eq.(\ref{eq:25}).  They assumed that the their "non-perturbative" eigenvalues comprised  the "double vacuum polarization"  and cancelled it against the contribution due to the scattering into intermediate excited states yielding effectively a zero change of the leading order Lamb shift.

\section{Conclusions}
 If the Lamb shift is taken as the dynamical QED effect due to the interaction of the muon with the vacuum polarization and not as a shift caused by the "Uehling potential" as an static external  potential, one gets agreement for the radius determined from the Lamb shift in muonic hydrogen with the combined electronic experiments. Since the relativistic calculations including the finite size effects of  Carroll et al. \cite{Carroll:2011rv} have to be redone realizing the new considerations of this paper we stick to the formula eq.\,(\ref{eq:1}) used by Pohl et al. \cite{Pohl10}. This means that all other corrections, in particular the "double vacuum polarization" and the finite size effect, are the same as used in that analysis. If we correct the Lamb shift of the point charge in eq.(\ref{eq:1}) to the value calculated in this paper, we arrive at a new value for the rms radius of the proton derived from the muon experiment:
 \begin{equation}
\langle r^2Ê\rangle^{1/2} = 0.87650(71)\,\textrm{fm}
\label{eq:27}
\end{equation}
where we have taken the recent 3rd Zemach moment from ref.\,\cite{Distler:2010zq} and included the estimated error of the point charge Lamb shift. This value is now in good agreement with the best electron scattering rms radius  0.879(8)\,fm \cite{Bernauer:2010wm}.
 
Again QED wins and there is no reason to fear a "chink in the armor" as  in ref. \cite{Flowers2010}. However, it is also true that "it must be repeated that the theory of relativistic effects and radiative corrections in bound sates is not yet in entirely satisfactory shape" \cite{Weinberg:1995mt}, p. 560.

\acknowledgments{I am indebted to my son Johannes Walcher for hints and remarks, to Marc Vanderhaeghen and Vladimir Pascalutsa for a critical discussion of the prefinal paper, and to Carl Carlson for providing me with his unpublished note.}
  
\bibliography{references}
 
\end{document}